\documentclass[pra,epsfigure,twocolumn,showpacs,preprintnumbers,amsmath,amssymb]{revtex4}

\usepackage{graphicx}
\usepackage{dcolumn}  
\usepackage{bm}  
\usepackage{epsfig}

\begin{document}
\setlength\arraycolsep{2pt}

\title{Casimir force in the presence of a medium}

\author{Fardin Kheirandish}
\email{fardin_kh@phys.ui.ac.ir} \affiliation{Department of
Physics, University of Isfahan, Isfahan 81746, Iran}
\affiliation{Quantum Optics Research Group, University of Isfahan,
Hezar Jarib Ave., Isfahan, Iran.}

\author{Morteza Soltani}
\email{msoltani@phys.ui.ac.ir} \affiliation{Department of Physics,
University of Isfahan, Isfahan 81746, Iran}

\author{Jalal Sarabadani}
\email{j.sarabadani@phys.ui.ac.ir} \affiliation{Department of
Physics, University of Isfahan, Isfahan 81746, Iran}

\begin{abstract}
In this article we investigate the Casimir effect in the presence
of a medium by quantizing the Electromagnetic (EM) field in the
presence of a magnetodielectric medium by using the path integral
formalism. For a given medium with definite electric and magnetic
susceptibilities, explicit expressions for the Casimir force are
obtained which are in agree with the original Casimir force
between two conducting parallel plates immersed in the quantum
electromagnetic vacuum.
\end{abstract}


\maketitle
\section{introduction}\label{intro}

One of the most remarkable and fundamentally important result of
the field quantization is the Casimir effect which is a force
arising from the change of the zero point energy caused by
imposing the boundary conditions (BC) \cite{Casimir}. This force
is the macroscopic aspect of the quantum electrodynamics that
provides a direct line between quantum field theory and the
macroscopic world. The original calculation of the Casimir force
between two perfectly conducting parallel plates immersed in the
quantum electromagnetic vacuum is based on the definition of the
Casimir energy in the presence and the absence of boundary
surfaces \cite{Casimir} that leads to an attractive observable
force
\begin{equation}
F=-\frac{\hbar c}{240}\frac{1}{H^4},
\end{equation}
between the plates, where $\hbar$ is the Planck constant, $c$ is
the speed of light and $H$ is the distance between the plates. In
other way one can consider this effect by evaluating the radiation
pressure on macroscopic objects \cite{Milonni}. As the magnitude
of the Casimir force is substantial at $H<100 ~{\text{nm}}$ this
effect is relevant in nano-technology
\cite{sriva,nanoscale1,nanoscale1b,s2} and should be take into
account to design and actuate microelectromechanical (MEMS)
systems. Moreover the possibility of transducing the energy from
the vacuum is investigated by means of MEMS
\cite{pinto,Cole,Forward}.

Many attempts have been focused on observing the Casimir force and
performing high-precision measurements during last few years
\cite{Lamoreaux,Mohideen,Harris,Bressi,Decca1,Decca2}. All these
experiments are in agree with the prediction of the Casimir
\cite{Casimir} within a few percents. These deviation from the
ideal force may due to temperature, roughness of surfaces and
finite dielectric constants that have been covered in a very
recent book entitled by Advanced in the Casimir Effect
\cite{Bordag_Book}.

\begin{figure}[b!]
\includegraphics[width=0.70\columnwidth]{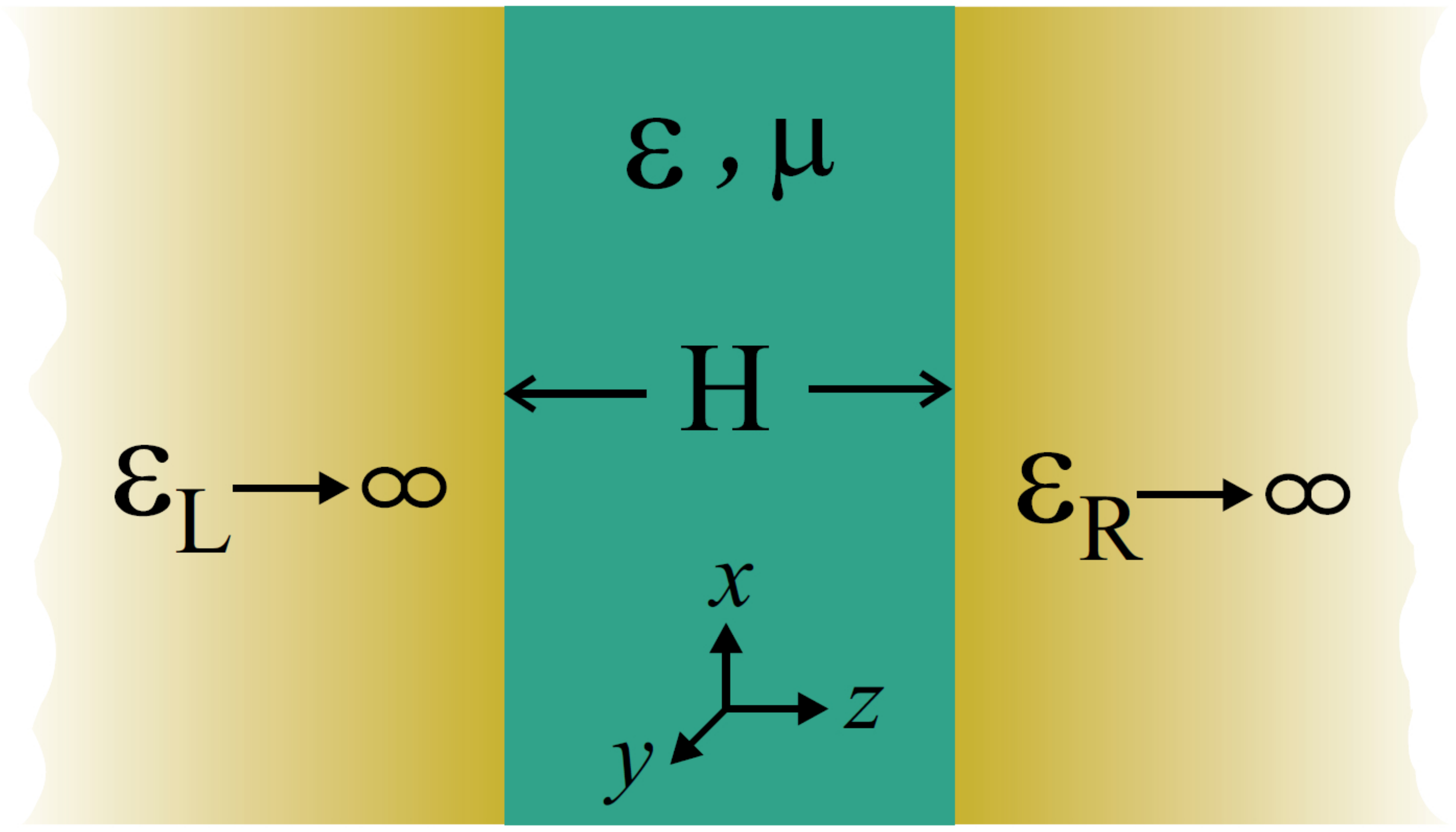}
\caption{This picture illustrates the schematic figure of the
system under consideration. A magnetodielectric medium
($\varepsilon, ~\! \mu$) is enclosed between two perfect parallel
conductors ($\varepsilon_{\text R}$ and $\varepsilon_{\text L}
\rightarrow \infty$). The distance between conductors is $H$, and
the $z$ direction is perpendicular to the surfaces of the media.}
\label{schematic-fig}
\end{figure}

In 90th decade, Golestanian and Kardar developed a path integral
approach to investigate the dynamic Casimir effect in the system
of two corrugated conducting plates surrounded by the quantum
vacuum \cite{Golestanian-PRL-1997,Golestanian-PRA-1998}. Emig and
his colleagues also used the path integral formalism to obtain
normal and lateral Casimir force between two sinusoidal corrugated
perfect conductor surfaces \cite{Emig-PRL-2001,Emig-PRA-2003}.
Later on, the exact mechanical response of the quantum vacuum to
the dynamic deformations of a cavity and the rate of dissipation
have been calculated by using path integral scheme \cite{jalal}.
This motivated us to investigate the Casimir effect in the
presence of a magnetodielectric medium by quantizing the
electromagnetic (EM) field using path integral formalism. Our
system contains of a magnetodielectric medium with permitivity
$\varepsilon$ and permeability $\mu$, enclosed by two
semi-infinite ideal metals ($ \varepsilon_{\text L} \rightarrow
\infty$ for the ideal metal in left-hand side and
$\varepsilon_{\text R} \rightarrow \infty$ for right-hand side
one) as depicted in Fig.~\!(\ref{schematic-fig}). We model the
magnetodielectric medium by a continuum of harmonic oscillators
(Hopfield Model) \cite{Fardin1,Fardin2,Fardin3}.

The outline of this paper is as follows: In Sec.\
\ref{path-scalar}, in order to introduce the scheme, we first
quantize the simple case of a scaler Klein-Gordon field in the
presence of a medium. Sec.\ \ref{force-scalar} is devoted to
obtain the Casimir force for the scalar filed in the presence of a
medium for different kinds of boundary conditions. In Sec.\
\ref{path-electromagnetic}, we develop our formalism to the case
of EM field in the presence of a magnetodielectric medium. Sec.\
\ref{force-electromagnetic} gives the Casimir force for the case
of EM field. Finally, the conclusions and outlooks are in the
Sec.\ \ref{conclusion}.

\section{Field quantization using path integrals}\label{path-scalar}

To illustrate the method and also for later convenience, before
considering the EM field in the presence of a medium, we consider
the simplest case i.e. a scaler massless field. In the next
section we will show that the Klein-Gordon field can be
corresponded to each polarization of the EM field. Therefore, let
us consider the following Lagrangian for  the total system
\begin{equation}\label{2}
{\cal {L}} = {\cal{L}} _{sys}+{\cal{L}} _{mat}+{\cal{L}}_ {int},
\end{equation}
where
\begin{equation}\label{3}
{\cal{L}}_{sys}  = \frac{1}{2}\,\partial ^\mu  \varphi\,\partial
_\mu \varphi,
\end{equation}
is the Lagrangian density of the massless Klein$-$Gordon filed.
The medium is modeled by a continuum of harmonic oscillators as
\cite{Huttner}
\begin{equation}\label{4}
{\cal{L}}_{mat}  = \int_0^\infty  d\omega\,(\frac{1}{2}\rho \dot
Y_\omega^2 - \frac{1}{2}\rho \omega ^2 Y_\omega^2),
\end{equation}
where $Y_\omega$ is an oscillator's field, $\rho$ is the density
of matter field and the interaction between the system and its
medium is defined by
\begin{equation}\label{5}
{\cal{L}}_{int}=\varphi \dot P,
\end{equation}
where
\begin{equation}\label{6}
P =\int d\omega \nu(\omega)Y_\omega.
\end{equation}
In the next section we will show that the quantity $P$ is in fact
the polarization field corresponding to the medium and the
interaction (\ref{5}) will become the electric-dipole interaction.

Generally a generating function is defined by \cite{Ryder}
\begin{equation}\label{7}
Z[J] = \int {\cal {D}} [\psi] \exp{\bigg\{\imath\int d^{n + 1}
x[{\cal L}\big(\psi (x)\big) + J(x)\psi (x)]\bigg\}},
\end{equation}
where $\psi$ is the scalar field and the different correlation
functions can be found by taking the repeated functional
derivatives with respect to the source field $J(x)$. The above
partition function is Gaussian since the integrand has quadratic
form with respect to the fields. To obtain the generating function
for the interacting fields, we first calculate the generating
function for the free fields
\begin{eqnarray}\label{8}
Z_0 \!\! && \!\! [J_{\varphi} ,J_\omega] =  \int
{\cal{D}}[\varphi]\prod_{\omega}{\cal{D}}[Y_\omega] \times
\nonumber\\
&& \hspace{-0.5cm} \times \exp \bigg\{ \imath \int d^{n+1} x \big[
{\cal{L}}_{sys} + {\cal{L}}_{mat} + J_{\varphi} \varphi + \int
d\omega J_\omega
Y_\omega \big ] \bigg\}. \nonumber\\
\end{eqnarray}
Using the $n$-dimensional version of Gauss's theorem we find
\begin{equation}\label{9}
\int d^{n+1} x\,\partial _\mu  \varphi\, \partial ^\mu  \varphi  =
-\int d^{n+1} x\,\varphi\,\Box\, \varphi,
\end{equation}
where $\Box$ is the d'Alemberian in $(n+1)$-dimensional space-time
and the integration by part
\begin{equation}\label{10}
 \int d^{n+1} x\,\,\dot Y_{\omega}\,\dot Y_{\omega} = -\int d^{n+1}
x\,Y_{\omega}\,\frac{{\partial ^2 }}{{\partial t^2 }}\,Y_{\omega}.
\end{equation}
the free generating function (\ref{8}) can be written as
\begin{eqnarray}\label{11}
Z_0 \!\! && \!\! [J_{\phi} ,J_\omega] = \int\prod_{\omega}{\cal
D}[Y_\omega]
{\cal D}[\varphi] \exp \bigg\{  - \frac{\imath}{2} \int d^{n + 1} x  \nonumber\\
&& \times \bigg[ \varphi (x)\Box\varphi (x) + \int d\omega
Y_\omega (x)(\frac{{\partial ^2 }}{{\partial t^2 }}+\rho
\omega^2)Y_\omega (x)
\nonumber\\
&& ~~~ + J_{\varphi} (x)\varphi (x) +\int d\omega J_\omega
(x)Y_\omega (x)\bigg]\bigg\}.
\end{eqnarray}
The integral in the equation (\ref{11}) can be easily calculated
from the field version of the quadratic integrals and the result
is
\begin{eqnarray}\label{12}
Z_0 \!\! && \!\! [J_{\varphi}, J_\omega] = \nonumber\\
&& \!\!\!\!\!\!\! = \exp \bigg\{ -\frac{\imath}{2} \int d^{n+1}x
\int d^{n+1} x' \!
\bigg[ J_{\varphi} (x) G_0 (x - x')J_{\varphi} (x') \nonumber\\
&& \hspace{+0.9cm} +  \int \!\! d\omega J_\omega (x) G_\omega  (x
- x') J_\omega(x') \bigg] \! \bigg\},
\end{eqnarray} where
$G_\omega (x - x')$ and $G_0 (x - x')$ are the propagators for
free fields and satisfy the following equations
\begin{equation}\label{13}
 \Box G_0 (x - x') = \delta (x - x'),
\end{equation}
\begin{equation}\label{14}
\{ \rho \frac{{\partial ^2 }}{{\partial t^2 }} + \rho \omega ^2 \}
G_\omega  (x - x') = \delta (x - x').
\end{equation}
We employ the Fourier transformation to solve the equations
(\ref{13}) and (\ref{14}). The solutions are
\begin{equation}\label{15}
G_0 (x - x') = \frac{1}{(2\pi)^{n+1}}\int d^n {\bf k}\,d\omega\,
\frac{{e^{\imath{\bf k}\cdot({\bf x} -{\bf x}') - \imath \omega (t
- t')} }}{{\omega ^2 - {\bf k}^2 }},
\end{equation}
and
\begin{equation}\label{16}
G_\omega  (x - x') = \frac{1}{(2\pi\rho)}\int d\omega'\,
\frac{{e^{ - \imath\omega' (t - t')} }}{{\omega^2-\omega'^2
}}\,\delta({\bf x}-{\bf x'}).
\end{equation}
Here the space component of the point $x\in {\cal R}^{n+1}$ is
indicated by the bold face ${\bf {x}}\in {\cal R}^{n}$ and the
time component by $t$ or $x_0\in {\cal R}$.

For further use we define
\begin{equation}\label{17}
J_p(z)=\int d\omega \nu(\omega) J_\omega(z),
\end{equation}
the generating function of the interacting fields can be written
in terms of the free generating function as \cite{Ryder}
\begin{eqnarray}\label{18}
Z[J_{\varphi},J_P] &=& Z^{ - 1} [0]e^{\imath \int d^{n+1}
z{\cal{L}}_{int} (\frac{\delta } {\delta J_\varphi
(z)},\frac{\delta }{\delta J_P(z)  })} Z_0 [J_{\varphi} ,J_\omega
] \nonumber\\
&& \hspace{-1.8cm} =  Z^{ - 1} [0] \sum\limits_{n = 0}^\infty
\frac{1}{n!}\bigg[ \imath\int d^{n+1}z\frac{\delta } {{\delta
J_\varphi(z) }} \cdot \frac{\partial }{{\partial z_0}}\frac{\delta
}{{\delta J_{\varphi} (z)  }} \bigg]^n \times \nonumber\\
&& \hspace{+0.6cm} \times Z_0 [J_{\varphi} ,J_\omega ],
\end{eqnarray}
where $Z [0]$ is the partition function of the free space. Thus
the Green's function of Klein-Gordon filed can be obtained via
\begin{equation}\label{19}
G_{\varphi\varphi}(x - y) = \imath \frac{{\delta ^2
Z[J_{\varphi},J_P]}}{{\delta J_{\varphi} (x)\delta J_{\varphi}
(y)}}\big |_{J_{\varphi},J_{\omega} = 0}.
\end{equation}
Combining the generating function (\ref{18}), the Green's function
(\ref{19}) and the definition of ${\cal L}_{int}$ (\ref{5}) yield
the following series for the Green's function
\begin{eqnarray}\label{20}
G_{\varphi\varphi}(x - x')& =& G_0 (x - x')\nonumber\\
&& \hspace{-2.4cm} + \!\! \int \!\! d\omega \!\! \int \!\! dx_1
dx_2 G_0 (x \! - \! x_1 \!) \nu ^2 (\omega )\frac{{\partial ^2
}}{{\partial t^2 }}G_\omega (x_1  \! - \! x_2 \! )G_0 (x_2 \!  -
\! x')
\nonumber\\
&& \hspace{-2.4cm} + \!\!\! \int \!\!\! d\omega \!\!\! \int \!\!\!
d\omega ' \!\!\!\! \int \!\!\! dx_1 dx_2 \!\!\! \int \!\! \! dx_3
dx_4 G_0 (x \! - \! x_1 \!)\nu ^2 (\omega )\frac{{\partial ^2
}}{{\partial t^2 }}G_\omega (x_1 \!\! - \! x_2 \! ) \!\! \times\nonumber\\
&& \hspace{-2.4cm}  \times G_0 (x_2 \! - \! x_3 \!)\nu ^2 (\omega
')\frac{{\partial ^2 }} {{\partial t^2 }}G_{\omega '} (x_3  \!-
\!x_4 \!)G_0 (x_4 \! - \! x') \! + \! \cdots  .
\end{eqnarray}
It is appropriate to use a general Green's function in
$(n+1)$-dimensional Fourier space that is
\begin{equation}\label{21}
G({\bf k},\omega )
 =\int e^{\imath {\bf k}\cdot{\bf x} - \imath \omega t}\, G(x)\,dt\,d^{n}\,{\bf x}.
\end{equation}
Therefore the Green's functions of the free fields $\varphi$ and
$Y_\omega$ in the Fourier space might be given by
\begin{equation}\label{22}
G_0 ({\bf k},\omega ) = \frac{1}{{{\bf k}^2  - \omega ^2 }},
\end{equation}
and
\begin{eqnarray}\label{23}
G_\omega  ({\bf k},\omega ') &=& \frac{1}{\rho}\int \frac{
d\omega'' dt\,d^3 {\bf x} ~{e^{-\imath\omega'' t} }}{{\omega^2 -
\omega ^{''2}  - \imath0^ + }}\delta ({\bf x})e^{\imath(\omega' t
- {\bf k}\cdot{\bf x})}
\nonumber \\
&=& \frac{1}{\rho}\frac{1}{\omega^2  - \omega^{'2} -\imath 0^{+}}
=: G_\omega (\omega '),
\end{eqnarray}
respectively. Since we are interested in retarded Green's
functions we have added $ - \imath0^+$ to the denominator of the
equation (\ref{23}). Since the reservoir field is assumed to be
homogeneous, the Green's function of the reservoir does not depend
on ${\bf k}$ in the above equation. Using the equations (\ref{22})
and (\ref{23}), $G_{\varphi\varphi}(x - x')$
can be written in the Fourier space as
\begin{eqnarray}\label{24}
G_{\varphi,\varphi}({\bf k},\omega )&& \nonumber\\
&& \hspace{-1.8cm} = G_0 ({\bf k},\omega ) \{1+\sum\limits_{n =
0}^\infty[\int d\omega'\omega^2\nu ^2 (\omega' )G_{\omega'}(\omega
) G_0({\bf k},\omega )]^n\}\nonumber\\
&& \hspace{-1.8cm} = \frac{G_0 ({\bf k},\omega )} {1 - \int
d\omega'\omega^2\nu ^2 (\omega' )G_{\omega'}(\omega)G_0 ({\bf
k},\omega )} \nonumber\\
&& \hspace{-1.8cm} = \frac{1}{{{\bf k}^2-\omega ^2-
\frac{1}{\rho}\int d\omega'\frac{\nu ^2 (\omega')\omega^2 }{\omega
'^2 -\omega ^2+ \imath 0^+ }}}.
\end{eqnarray}
This Green's function can also be obtained directly from the
Heisenberg equations of motion. By direct substitution we can show
that the Green's function $G_{\varphi\varphi}(x-y)$ satisfies the
equation
\begin{eqnarray}\label{25}
&& \!\!\!\!\! \Box G_{\varphi\varphi} ({\bf x} - {\bf x}',t - t') \nonumber\\
&& \!\!\!\!\! - \frac{\partial }{{\partial t}} \!\!
\int_{-\infty}^{t} \!\!\!\!\! dt'' \! \chi (t \! - \! t''
)\frac{\partial }{{\partial t''}}G_{\varphi\varphi} ({\bf x} \!-\!
{\bf x}',t'' \! \!- \! t') \!=\! \delta ({\bf x} \!-\! {\bf x}',t
\! - \! t'),
\nonumber\\
\end{eqnarray}
which is the motion equation of dissipation field with
susceptibility of the medium $\chi(t)$, with the following Fourier
transform
\begin{equation}\label{26}
\chi(\omega ) = \frac{1}{\rho}\int d\omega '\frac{{\nu ^2 (\omega
')}}{{\omega ^{'2} - \omega ^2  + \imath 0^+ }}.
\end{equation}
From the equations (\ref{24}) and (\ref{26}) it is clear that the
modified Green's function $G_{\varphi,\varphi}({\bf k},\omega )$
can be obtained from the free field Green's function $G_0 (x-x')$
in Eq.~\!(\ref{15}) simply by replacing $\omega^2$ with
$\varepsilon(\omega)\omega^2$, where
$\varepsilon(\omega):=1+\tilde{\chi}(\omega)$. It can be easily
shown that this susceptibility satisfies the Kramers-Kronig
relations as expected.

By the same technique we can obtain correlation between the
polarization field and the Klein-Gordon field. To this end we
define $G_{\varphi,P} $ as
\begin{equation}\label{27}
G_{\varphi,P}=\frac{\delta^2 Z[J_\varphi, J_P]}{\delta J_\varphi
\delta J_P},
\end{equation}
that can be obtained via direct calculation similar to the
procedure that ended to $G_{\varphi,\varphi}({\bf k},\omega )$ as
\begin{eqnarray}\label{28}
&&G_{\varphi,P}(x - x') = \int dx_1G_0 (x - x_1)\frac{\partial}{\partial z_0}G_{0P}(x_1-x')\nonumber\\
&& +  \int dx_1 dx_2 G_{0P} (x - x_1 ) \frac{{\partial ^2
}}{{\partial t^2 }}G_0  (x_1  - x_2 )G_{0P} (x_2 - x')
\nonumber\\
&& + \int dx_1 dx_2 \int dx_3 dx_4 G_{0P} (x - x_1
)\frac{{\partial ^2 }}{{\partial t^2 }}G_0 (x_1  - x_2
) \times \nonumber\\
&& \hspace{+0.0cm} \times G_{0P}(x_2 - x_3 )\frac{{\partial ^2 }}
{{\partial t^2 }}G_0 (x_3 \! - \! x_4 )G_{0P} (x_4 \! - \! x') +
\cdots,
\end{eqnarray}
where
\begin{equation}\label{29}
G_{0P}(x-x')=\int d\omega \nu^2(\omega)G_\omega(x-x').
\end{equation}
If we again write $G_{\varphi,P}(x - x')$ in the Fourier space we
find
\begin{eqnarray}\label{30}
G_{\varphi,P}({\bf k},\omega )&=& \imath \omega G_0 ({\bf
k},\omega )
G_{0P}({\bf k},\omega ) \nonumber\\
&& \hspace{-0.6cm} \times \big\{1+\sum\limits_{n = 0}^\infty[\int
d\omega'\omega^2\nu ^2 (\omega' )G_{\omega'}(\omega ) G_0({\bf
k},\omega )]^n \big\}\nonumber\\
&=& \frac{G_{0P} ({\bf k},\omega )  G_0 ({\bf k},\omega )} {1 -
\int d\omega'\omega^2\nu ^2 (\omega' )G_{\omega'}(\omega)G_0
({\bf k},\omega )}\nonumber\\
&=& \frac{G_{0P} ({\bf k},\omega )}{{{\bf k}^2-\omega ^2-
\frac{1}{\rho}\int d\omega'\frac{\nu ^2 (\omega')\omega^2 }{\omega
'^2 -\omega ^2+ \imath 0^+ }}}.
\end{eqnarray}
Comparing Eqs.~\!(\ref{16}) and (\ref{29}), yields $G_{0P} ({\bf
k}, \omega)=\int d\omega\frac{\nu^2(\omega')}{\omega'^2-\omega^2+
\imath 0^+}=\tilde{\chi}(\omega)$, consequently
$G_{\varphi,P}({\bf k},\omega )$ can be rewritten as
\begin{equation}\label{31}
G_{\varphi,P} ({\bf k},\omega )=\imath
\omega\chi(\omega)G_{\varphi,\varphi} ({\bf k},\omega).
\end{equation}
The other important correlation function is $G_{P,P}({x-x'})$
which is defined via generating function $Z[J_\varphi,J_P]$ as
\begin{equation}\label{32}
G_{P,P}({x-x'})=\frac{\delta^2Z[J_\varphi,J_P]}{\delta J_P\delta
J_P}.
\end{equation}
By straightforward calculations one can obtain $G_{P,P}({\bf
k},\omega)$ as
\begin{equation}\label{33}
G_{P,P}({\bf
k},\omega)=\frac{\nu^2(\omega)}{\omega}+\omega^2\chi^2(\omega)G_{\varphi\varphi}({\bf
k},\omega).
\end{equation}
The imaginary part of the response function can be read from
Eq.~\!(\ref{26}) as $\frac{\nu^2(\omega)}{\omega}=Im\chi(\omega)$.
Here to illustrate the validity of our results, i.e.
Eq.~\!(\ref{31}-\ref{33}), we compare them with the results of the
other methods of field quantization. In other conventional methods
of phenomenological field quantization \cite{Matloob}, the fields
can be divided into positive ($+$) and negative ($-$) frequencies
parts which satisfy the constitutive relation
\begin{equation}\label{34}
\hat{P}^{\pm}({\bf k},\omega)= \pm \imath
\omega\chi(\omega)\hat{\varphi}^{\pm}({\bf
k},\omega)+\hat{P}_N({\bf k}^{\pm},\omega),
\end{equation}
after quantization of the fields, where by $\hat{}$ we mean
operator and the operator with positive frequency is the Hermitian
conjugate of the negative one. $\hat{P}_N^{\pm}$ is the noise part
of the polarization field that according to the
fluctuation-dissipation theorem \cite{Matloob, Fardin3} satisfies
%
%
\begin{equation}\label{35}
[ \hat{P}_N^+ ({{\bf x},\omega}) , \hat{P}_N^{-} ({\bf
x},\omega)]= \pi Im\chi(\omega)\delta({\bf x}-{\bf x}'),
\end{equation}
which $[\cdots,\cdots]$ denotes the commutator of two operators.
If we use the Eqs.~\!(\ref{34}) and (\ref{35}) to obtain the
Green's functions, we achieve the same results as the
Eqs.~\!(\ref{31}-\ref{33}). This shows the validity of our path
integral quantization.

According to the definition of the Green's functions, after
integrating over gaussian fields one can read the generating
function (\ref{18}) as
\begin{eqnarray}\label{36}
Z[J_\varphi,J_P]&&=exp\bigg\{ \imath \int dx \int dx' \big[
J_\varphi G_{\varphi,\varphi}(x-x')J_\varphi \nonumber\\
&& +J_\varphi G_{P,\varphi}(x-x')J_P+J_PG_{P,P}(x-x')J_P \big]
\bigg\}. \nonumber\\
\end{eqnarray}

\section{Calculating the Casimir force}\label{force-scalar}

\subsection{General formalism}
In this section we briefly review the path integral technique to
calculate the Casimir force. Let us consider two conducting plates
faced each other at the distance $H$ and embedded in an arbitrary
medium. The field $\varphi$ satisfies the Dirichlet
\begin{equation}\label{37}
\varphi (X_\alpha) = 0,
\end{equation}
or Neumann
\begin{equation}\label{38}
{\partial_n}\varphi (X_\alpha) = 0,
\end{equation}
 boundary conditions on surface, where $X_\alpha, (\alpha=1,2)$ is an
arbitrary point on the $\alpha$th conducting plate.  To obtain the
partition function from the Lagrangian we use the Wick's rotation,
$(t\rightarrow \imath\tau)$ and change the signature of the
space-time from Minkowski to Euclidean. The Diriclet or Neumann
boundary conditions can be taken into account using the auxiliary
fields $\psi _\alpha (X_\alpha )$ \cite{jalal}
\begin{equation}\label{38}
\delta \big( \varphi (X_\alpha  ) \big) = \int {\cal
{D}}[\psi_\alpha (X_\alpha )]e^{\imath\int d X_\alpha \psi
(X_\alpha  )\varphi (X_\alpha  )}.
\end{equation}
and
\begin{equation}\label{39}
\delta \big({\partial_n}\varphi (X_\alpha  ) \big) = \int {\cal
{D}}[\psi_\alpha (X_\alpha )]e^{- \imath\int d X_\alpha
{\partial_n}\psi (X_\alpha )\varphi (X_\alpha  )}.
\end{equation}
After Wick's rotation the Dirichlet and Neumann partition
functions can be cast into the form
\begin{equation}\label{jalal}
 Z_D = Z_0^{ - 1} \int {\cal {D}}[\varphi] \prod\limits_{^{a = 1} }^2
  {\cal D}[\psi_\alpha (X_\alpha  )])e^{S_D[\varphi ]},
\end{equation}
and
\begin{equation}\label{41}
 Z_N = Z_0^{ - 1} \int {\cal {D}}[\varphi] \prod\limits_{^{a = 1} }^2
  {\cal D}[\psi_\alpha (X_\alpha  )])e^{S_N[\varphi ]}
\end{equation}
respectively, where $Z_0$ is the partition function of the free
space, and
\begin{eqnarray}\label{42}
&&S_D  [\varphi ] = \int d^{(n+1)} x  \nonumber\\
&&  \times \big\{ {\cal L} \big( \varphi(x) \big)+ \varphi (x)\!\!
\sum\limits_{\alpha = 1}^2 \! \int \!\! d^{(n)} X\delta (X \!\! -
\!\! X_\alpha )\psi _\alpha  (x) \big\}.
\end{eqnarray}
and
\begin{eqnarray}\label{43}
S_N [\varphi ] = \int d^{(n+1)} x && \nonumber\\
&& \hspace{-3.8cm} \times \{{\cal L} (\varphi(x))+ \varphi
(x)\sum\limits_{\alpha = 1}^2 \int d^{(n)} X\delta (X - X_\alpha
)\partial_n\psi _\alpha (x)\}.
\end{eqnarray}
Using the same procedure of Ref.~\!(\cite{jalal}) the parttion
functions for the Dirichlet and Neumann BC can be read
\begin{equation}\label{52}
Z_D  = \frac{1}{{\sqrt {\det \Gamma_D (x,y,H)} }},
\end{equation}
and
\begin{equation}\label{53}
Z_N  = \frac{1}{{\sqrt {\det \Gamma_N (x,y,H)} }},
\end{equation}
where
\begin{equation}\label{54}
\Gamma_D (x,y,H) =\bigg[\begin{array}{*{20}c}
   {{\cal G}(x - y,0)} & {{\cal G}(x - y,H)}  \\
   {{\cal G}(x - y,H)} & {{\cal G}(x - y,0)}  \\
\end{array}\bigg],
\end{equation}
and
\begin{equation}\label{55}
\Gamma_N (x,y,H) =\bigg[\begin{array}{*{20}c}
-\partial^2_z{{\cal G}(x - y,0)} & -\partial^2_z{{\cal G}(x - y,H)}  \\
-\partial^2_z{{\cal G}(x - y,H)} & -\partial^2_z{{\cal G}(x - y,0)}  \\
\end{array}\bigg],
\end{equation}
where ${\cal G}$ is the Green's function of the fields after wick
rotation. We define the effective action as
\begin{equation}\label{56}
S_{eff}  =- \imath \ln Z (H),
\end{equation}
where $\ln Z (H)$ can be either for the Dirichlet or Neumann BC,
in order to calculate the Casimir force by applying derivative
with respect to the distance between the plates
\begin{equation}\label{57}
F = \frac{{\partial S_{eff} (H)}}{{\partial H}}.
\end{equation}
It is easy to show that the contribution of the Dirichlet is the
same as that of the Neumann BC to the Casimir energy and hence the
Casimir force in the presence of a isotropic and homogenous medium
like \cite{Emig-PRA-2003}. So that in the next sections we treat
only the Dirichlet BC.

\subsection{Casimir force for different boundary
conditions}\label{Casimir-Different-BC-Scalar}

In this section we would like to obtain the Casimir force in the
presence of an absorptive medium. Before we obtain the Casimir
force for interacting fields, we investigate the possibility of
the existence of the Casimir force due to the matter field alone.
Using the Lagrangian (\ref{4}) and expression for the effective
action (\ref{56}) we find the $\Gamma_\omega $ tensor as
\begin{equation}\label{58}
\Gamma_{\omega} (x,y,H) =\bigg[\begin{array}{*{20}c}
   {{\cal G}_\omega(x - y,0)} & {{\cal G}_\omega(x - y,H)}  \\
   {{\cal G}_\omega(x - y,H)} & {{\cal G}_\omega(x - y,0)}  \\
\end{array}\bigg].
\end{equation}
But since ${\cal G}_\omega(x-y,H)=0$ for this situation, the
noninteracting matter field alone, does not lead to any modified
Casimir force. This result is clear since we model the matter
field by the Hopfield model ******[HOPFIELD's REFERENCE]******.
This model is based on an independent set of harmonic oscillators
and imposing any condition on one of these oscillators does not
affect the others, and hence we do not expect any Casimir effect.

For a dissipative field $\varphi$ (\ref{25}) we may consider three
different boundary conditions,

{\bf i}) Imposing the boundary condition on the Klein-Gordon
field: for this case the $\Gamma$ tensor can be read
\begin{equation}\label{59}
\Gamma_{\varphi\varphi} (x,y,H) =\bigg[\begin{array}{*{20}c}
{{\cal G}_{\varphi,\varphi}(x - y,0)} & {{\cal G}_{\varphi,\varphi}(x - y,H)}  \\
{{\cal G}_{\varphi,\varphi}(x - y,H)} & {{\cal G}_{\varphi,\varphi}(x - y,0)}  \\
\end{array}\bigg].
\end{equation}
Since $\Gamma_{\varphi \varphi}$ is diagonal in the Fourier space,
to obtain the Casimir force we proceed in this space. The Fourier
transformation of $G_{\varphi \varphi} (x -y,H)$ is
\begin{eqnarray}\label{60}
{\cal G}_{\varphi \varphi}(p,q,H) &=& \int dx d y e^{\imath p.x
+\imath q.y} G_{\varphi \varphi} (x -y,H)\nonumber\\
&=&\frac{{e^{ - n(p_0 )|p_0| h} }}{2n(p_0 )|p_0|} (2
\pi)^3\delta(p+q),
\nonumber\\
\end{eqnarray}
%
where $p= (p_0, {\bf p})$, $\bf p$ is a vector parallel to the
conductor, $p_0$ the temporal component of the $p$, $n(p_0)=
\sqrt{\bar{\varepsilon}(p_0)}$ and
$\bar{\varepsilon}(p_0)=\varepsilon (\imath \omega)$. Thus for the
case {\bf i} the Casimir force is
\begin{equation}\label{61}
F_{\bf i} = - \int \frac{d^3 p }{(2\pi )^3
}[\frac{E(p)}{e^{2E(p)h} - 1}],
\end{equation}
where $E(p)=[n^2 (p_0 )p_0^2  + {\bf p}^2]^{1/2}$. In the absence
of the medium between the conductors, $n(p_0)=1$, we recover the
original Casimir force between two plates immersed in the quantum
vacuum of a scalar field
\begin{equation}\label{62}
F_{\bf i}= - \int\frac{d^3p}{(2\pi)^3}\frac{p^2}{e^{2|p| H}-1}= -
\frac{\pi^2}{480 H^4},
\end{equation}
and for a non absorptive medium with the susceptibility
$\chi(t)=\chi_0 \delta(t)$, we find the modified Casimir force as
\begin{equation}\label{63}
F_{\bf i}=\frac{1}{n}F_{\text{Vac}}.
\end{equation}
The above relation is fully in agree with the result of the
Lifshitz theory of fluctuation-induced force bewteen media
\cite{Lifshitz},\cite{Milonni-Book}. The equation (\ref{61}) is
interesting since it is the reminiscent of the Bose-Einstein
distribution. In fact, $E(p)$ can be interpreted as the force
density due to the bosons in the state $p$.

{\bf ii}) Imposing the boundary condition on the polarization
field: in this case $\Gamma_{PP}$ tensor is
\begin{equation}\label{64}
\Gamma_{PP} (x,y,H) =\bigg[\begin{array}{*{20}c}
{{\cal G}_{P,P}(x - y,0)} & {{\cal G}_{P,P}(x - y,H)}  \\
{{\cal G}_{P,P}(x - y,H)} & {{\cal G}_{P,P}(x - y,0)}  \\
\end{array}\bigg].
\end{equation}
Here the Casimir force is
\begin{equation}\label{65}
F_{\bf ii} = - \int \frac{d^3 p }{(2\pi )^3
}[\bar{\chi}^2(p_0)\frac{E(p)}{\alpha e^{2E(p)H} - 1}],
\end{equation}
where $\alpha =E(p)Im\bar{\chi}(p_0)+\bar{\chi}^2(p_0)$. Although
the noise operators do not have any spatial correlation but their
presence on the surface can affect the Casimir force and decrease
it due to polarization.

{\bf iii}) Imposing the boundary condition on the both of
polarization and Klein-Gordon fields: we can easily show that
$\Gamma_{\varphi\varphi,PP} (x,y,H)$ is a $8 \times 8$ tensor with
the form of
\begin{eqnarray}\label{65}
\Gamma_{\varphi\varphi,PP} (p,q,H) &&
\nonumber\\
&& \hspace{-2.6cm} =\bigg[\begin{array}{*{20}c}
{\!\Gamma_{\!\varphi\varphi} (p,q,\!H\!) }&{q_0 \bar{\chi}(q_0)\Gamma_{\! \varphi\varphi} (p,q,\!H\!) \!}  \\
{\!\!q_0 \bar{\chi}(q_0)\Gamma_{\!\varphi\varphi} (p,q,\!H\!)} &
{~\!q_0^2 \chi^2(q_0) \Gamma_{\! \varphi\varphi} (p,q,\!H\!) \!+\! Im \bar{\chi} (q_0) {\bf I}\!\!}  \\
\end{array}\bigg], \nonumber\\
\end{eqnarray}
where ${\bf I}$ is the $2\times 2$ unit matrix multiplied by $(2
\pi)^3 \delta (p+q)$.
%
It can be easily shown in this case that the Casimir force will be
the same as that of imposing the boundary condition only on the
Klein-Gordon field i.e. case {\bf i}. We can interpret this
situation with the aid of Eq.~\!(\ref{34}). According to this
relation if the BC is imposed only on the Klein-Gordon field then
$P_N^{\pm}$ is not zero, and if the BCs are imposed on the both of
polarization and Klein-Gordon fields then the BC will be imposed
on $P_N^{\pm}$ automatically which according to the begining of
the Sec.\ \ref{Casimir-Different-BC-Scalar} does not lead to aany
Casimir effect.

The physical interpretation of the first and third condition is
obvious. These conditions arise when we want to calculate the
Casimir force between two perfect conductors that enclose a
medium. But the second BC can happen when we want to consider the
system of containing two dielectric slabs. This kind of boundary
condition leads to the Casimir force between two dielectric slabs
which is under consideration. Here we only consider the boundary
conditions in cases {\bf i} and {\bf iii}, and we shall treat the
case {\bf ii} in elsewhere \cite{Fardin4}.

\section{Electromagnetic field quantization in the presence of a magnetodielectric medium}\label{path-electromagnetic}
In this section we develop the formalism to EM field in the
presence of a magnetodielectric medium \cite{Fardin3}. The
Lagrangian of EM field in the presence of a medium can be written
as
\begin{equation}\label{67}
{\cal L}={\cal L}_{EM}+{\cal L}_{1mat}+{\cal L}_{2mat}+{\cal
L}_{int}
\end{equation}
where ${\cal L}_{EM}$ is
\begin{equation}\label{68}
{\cal L}_{EM}=\frac{\varepsilon_0{\bf E}^2}{2}-\frac{{\bf
B}^2}{2\mu_0},
\end{equation}
where ${\bf E}$ and ${\bf B}$ are the electric and magnetic
fields. They can be written in terms of scalar and vector
potentials $U$ and ${\bf A}$ respectively as ${\bf E}=\dot {\bf
A}-\nabla  U$ and ${\bf B}=\nabla\times{\bf A}$. In this work we
use the Coulomb gauge $\nabla\cdot {\bf A}=0$, i.e. ${\bf A}$ is a
transverse field. ${\cal L}_{1mat}$ and ${\cal L}_{2mat}$ reffer
to the polarization and magnetization of the medium respectively
and can be written as
\begin{equation}\label{69}
{\cal L}_{{\bf i}mat}=\int_0^\infty d\omega \big(
\frac{1}{2}{\dot{\bf X}_{{\bf
i}\omega}^2}+\frac{1}{2}\omega^2{{\bf X}_{{\bf i}\omega}^2} \big),
\end{equation}
where ${\bf X}_{{\bf i}\omega}$ is an oscillator's vector field.
The interaction part of the Lagrangian is
\begin{equation}\label{71}
{\cal L}_{int}={\bf A}\cdot \dot{\bf P}- \nabla U . {\bf P}
+\nabla\times{\bf A}\cdot{\bf M},
\end{equation}
where ${\bf P}$ and ${\bf M}$ are polarization and magnetization
of the medium defined by
\begin{eqnarray}\label{70}
{\bf P}&=&\int_0^\infty\nu_1 (\omega){\bf X}_{1\omega},\nonumber\\
{\bf M}&=&\int_0^\infty\nu_2 (\omega){\bf X}_{2\omega}.
\end{eqnarray}
The Euler-Lagrange equations lead us to the fact that $\dot U$ is
not an independent dynamical variable. To obtain $U$ in terms of
the other independent dynamical variables, it is appropriate to
write the fields in Fourier space where the longitudinal and
transverse parts of a field can be separated using the unit
vectors ${\bf e}_3({\bf k})=\hat{\bf k}$ and ${\bf e}_\lambda({\bf
k})$ ($\lambda=1,2$) respectively. ${\bf e}_1({\bf k})$ and ${\bf
e}_2({\bf k})$ are perpendicular to each other and $\hat{{\bf
k}}$. Using these unit vectors the scalar potential can be written
in terms of the longitudinal part of the matter field as (In what
follows we indicate the fields in Fourier space by a $\tilde{}$
over them.)
\begin{equation}\label{72}
\tilde{U}= \imath \frac{\tilde{X}^\parallel_1}{\varepsilon_0 |{\bf
k}|}.
\end{equation}
By applying this recent relation to the Eq.~\!(\ref{67}), we can
easily show that the longitudinal part of the electromagnetic
field only changes the longitudinal component of ${\cal L}_{1mat}$
as
\begin{equation}\label{73}
\tilde{{\cal L}}^\parallel_{1mat}=\int_0^\infty d\omega \big(
\frac{1}{2}{\dot{\tilde{ X}}^{\parallel 2}_{1
\omega}}+\frac{1}{2}\omega'^2{\tilde{ X}^{\parallel 2}_{1 \omega}}
\big)
\end{equation}
where $\omega'=\sqrt{\omega^2+\omega^2_c}$ and
$\omega^2_c=\frac{\nu^2(\omega)}{\varepsilon_0}$. It is worth
noting that only the transverse parts of Lagrangians have
contribution to the Casimir effect and the longitudinal part does
not lead to any Casimir force. Therefore we just consider the
transverse part of the Lagrangian which in the Fourier space is
\begin{equation}\label{74}
L^\perp= \int'{d^3{\bf k}\tilde{\cal L}^\perp_{em}+\sum_{{\bf
i}=1,2}(\tilde{\cal L}^\perp_{{\bf i},mat}+ \tilde{\cal
L}^\perp_{{\bf i},int})},
\end{equation}
where
\begin{equation}\label{75}
\tilde{{\cal L}}^\perp_{em}=\varepsilon_0( \dot{\tilde{{\bf A}}
}^2-c^2{\bf \tilde B}^2),
\end{equation}
\begin{equation}\label{76}
\tilde{{\cal L}}^\perp_{{\bf i},mat}=\int d \omega \big( \rho
\dot{\tilde{\textbf{X}}}_{\bf i}^{\perp2}-\rho \omega^{2}{\bf
\tilde X}_{\bf i}^{\perp2} \big) ,
\end{equation}
and $\tilde{\cal L}^\perp_{1,int}$ and $\tilde{\cal
L}^\perp_{2,int}$ refer to the interaction of the polarization and
magnetization fields with the electromagnetic field respectively
which are
\begin{equation}\label{77}
\tilde{\cal L}^\perp_{1,int}= \int_0^\infty \!\!  d\omega \nu_1 (
\omega )
\tilde{\textbf{A}}\cdot\dot{\tilde{\textbf{X}}}_1^\perp=\sum_{\lambda=1}^2\int_0^\infty
\!\! d\omega \nu_1 ( \omega ) \tilde{A}_\lambda \dot{\tilde{X}}_{1
\lambda},
\end{equation}
and
\begin{eqnarray}\label{78}
\tilde{\cal L}^\perp_{2,int} && =  \int_0^\infty d\omega \nu_2 (
\omega)
\textbf{k}\times\tilde{\bf{A}}\cdot\dot{\tilde{\textbf{X}}}_2^\perp
\nonumber\\
&&=\sum_{\lambda,\lambda'=1}^2\int_0^\infty d\omega \nu_2 ( \omega
) |{\bf k}| \tilde{A}_{\lambda}
{\tilde{\text{X}}}_{2\lambda'}\epsilon_{\lambda\lambda'},
\end{eqnarray}
where $\epsilon_{\lambda\lambda'}$ is the antisymmetric tensor.

To obtain the Casimir force between two plates with axial symmetry
due to the transverse part of the Lagrangian, the filed modes can
be divided into TM and TE modes
\cite{Emig-PRL-2001,Emig-PRA-2003}. It can be shown that for our
case as we consider the electromagnetic field in the presence of a
homogenous, isotropic and flat magnetodielectric medium, similar
to \cite{Emig-PRL-2001,Emig-PRA-2003} again the field can be
divided into TM and TE modes which refer to the Dirichlet and
Neumann BC respectively. As for the situation under study here for
TM and TE modes the Casimir force is the same, therefore here we
consider only the TM mode. According to the above discussion the
full Lagrangian can be rewritten as
\begin{equation}\label{79}
{\cal {L}} = {\cal{L}} _{sys}+{\cal{L}} _{mat}+{\cal{L}}_ {int},
\end{equation}
where
\begin{equation}\label{80}
{\cal{L}}_{sys}  = \frac{1}{2}\,\partial ^\mu  \varphi\,\partial
_\mu \varphi,
\end{equation}
is the Lagrangian density of a massless Klein$-$Gordon filed and
\begin{equation}\label{81}
{\cal{L}}_{mat}  =\sum_{{\bf i}=1}^2 \int_0^\infty
d\omega\,(\frac{1}{2}\rho \dot X_{{\bf i}\omega}^2 -
\frac{1}{2}\rho \omega ^2 X_{ {\bf i}\omega}^2).
\end{equation}
The interaction term is defined by
\begin{equation}\label{82}
{\cal{L}}_{int}=\varphi \dot P+|\nabla \varphi|M,
\end{equation}
where $P=\int d\omega\nu_1(\omega)X_{1\omega}$ and $M=\int
d\omega\nu_2(\omega)X_{2\omega}$. It can be seen that the
Lagrangian (\ref{79}) is similar to the Lagrangian (\ref{2}). In
fact these Lagrangians are the same if we take $\nu_2(\omega)=0$.
This is the reason why we called $P$ as a polarization field and
the results obtained in the Sec.\ \ref{path-scalar} can be used
for a polarizable medium.

By mixing (\ref{18}) and (\ref{82}), and use the same procedure in
the Sec.\ \ref{path-scalar}, after some manipulations we obtain
the Green's function as
\begin{eqnarray}\label{83}
G_{\varphi,\varphi}({\bf k},\omega)=\frac{1}{{\bf k} ^2 (1 - \chi
_m (\omega )) - \omega ^2 (1 + \chi _e (\omega))},
\end{eqnarray}
which is the Green's function of EM field in the presence of
magnetodielectric medium with the electric and magnetic
susceptibilities
\begin{equation}\label{84}
\chi_e (\omega ) \equiv\int_{ - \infty }^{+\infty}d\omega
'\frac{\nu_1^2(\omega ')}{{\omega - \omega ' +\imath 0^+}},
\end{equation}
and
\begin{equation}\label{85}
\chi_m (\omega )\equiv \int_{ - \infty }^{+\infty}d\omega
'\frac{\nu^2_2(\omega ')}{{\omega - \omega ' +\imath 0^+}},
\end{equation}
respectively. These susceptibilities satisfy the Kramers -Kronig
relations as expected. The other Green's functions or correlation
functions are
\begin{equation}\label{86}
G_{\varphi,P}({\bf k},\omega)=\imath \omega
\chi_e(\omega)G_{\varphi,\varphi}({\bf k},\omega),
\end{equation}
\begin{equation}\label{87}
G_{\varphi,M}({\bf k},\omega)=\imath |{\bf k}|
\omega\chi_m(\omega)G_{\varphi,\varphi}({\bf k},\omega),
\end{equation}
\begin{equation}\label{88}
G_{P,P}({\bf k},\omega)= \frac{\nu^2(\omega)}{\omega}+\omega^2
\chi_e^2(\omega)G_{\varphi \varphi}({\bf k},\omega),
\end{equation}
\begin{equation}\label{89}
G_{M,M}({\bf k},\omega)=\frac{\nu^2(\omega)}{\omega}+|{\bf
k}|^2\chi_m^2(\omega)G_{\varphi \varphi}({\bf k},\omega).
\end{equation}
The first terms in the right hand side of Eqs.~\!(\ref{88}) and
(\ref{89}) are related to the noise operators of the system which
satisfy the fluctuation-dissipation theorem, like (\ref{35}).
%
%
Consequently the generating function is
\begin{eqnarray}\label{90}
Z[J_\varphi,J_P]&=&\exp\bigg\{\imath \int dx \int dx' \bigg[
J_\varphi(x)
G_{\varphi,\varphi}(x-x')J_\varphi(x)\nonumber\\
&& \hspace{-2cm} + J_\varphi(x)
G_{P,\varphi}(x-x')J_P(x)+J_\varphi(x)
G_{P,\varphi}(x-x')J_P(x)\bigg]\bigg\}. \nonumber\\
\end{eqnarray}
This relation can be used to obtain the Casimir force in the next
section.
%
\section{Calculating the Casimir force for EM field in the presence of
a magnetodielectric medium}\label{force-electromagnetic}
%
%
Here again we consider three different boundary conditions similar
to the Sec.\ \ref{force-scalar} varies kinds of the fields.

{\bf i}) Imposing the boundary condition on EM field: if we impose
the Dirichlet BC on EM field the partition function becomes
\begin{equation}\label{91}
Z_D  = \frac{1}{{\sqrt {\det \Gamma_{\varphi \varphi} (x,y,H)} }},
\end{equation}
where
\begin{equation}\label{92}
\Gamma_{\varphi \varphi} (x,y,h H) =\bigg[\begin{array}{*{20}c}
{{\cal G}_{\varphi,\varphi}(x - y,0)} & {{\cal G}_{\varphi,\varphi}(x - y,H)}  \\
{{\cal G}_{\varphi,\varphi}(x - y,H)} & {{\cal G}_{\varphi,\varphi}(x - y,0)}  \\
\end{array}\bigg],
\end{equation}
and ${{\cal G}_{\varphi,\varphi}(x - y,h)}$ is the Green's
function (\ref{83}) with imaginary time.

To calculate $Z_D$, we invoke the Fourier space where
$\Gamma_{\varphi \varphi} (x,y,h H)$ is diagonal and its elements
have the form
\begin{eqnarray}\label{93}
&&{\cal G}_{\varphi , \varphi}(p,q,H)  \nonumber\\
&& = \bar{\mu}(p_0)\frac{e^{-\sqrt{n^2(p_0 )p_0^2 + {\bf p}^2}
~\!H}}{2\sqrt{n^2(p_0)
p_0^2+ {\bf p}^2}} (2 \pi)^3 \delta(p+q),
\end{eqnarray}
%
%
where here $n (p_0)=\sqrt{\bar{\mu}(p_0)\bar{\varepsilon}( p_0)}$
and $\mu(\omega)=\frac{1}{1-\chi_m(\omega)}$. Finally the Casimir
force is obtained as
\begin{equation}\label{94}
F = \int \frac{d^3 p }{(2\pi )^3 }[\frac{E(p)}{e^{2E(p)h} - 1}],
\end{equation}
where $E(p)=\sqrt{n^2(p_0 )p_0^2+{\bf p}^2}$. This equation is
like the equation (\ref{61}) the only difference is in the
definition of $n(p_0)$. If we impose the Neumann BC on the EM
field we will achieve the same result as that of the Dirichlet BC.

{\bf ii}) We can impose the BC on polarization and magnetization
fields. As we mentioned after the equation (\ref{65}), this
situation does not appear in our problem since we consider a
magnetodielectric medium surrounded by two perfect conductors
which lead to the BC on EM field. But for two magnetodielectric
slabs this kind of BC may appear which may be dealt with elsewhere
\cite{Fardin4}.

{\bf iii}) We can impose the boundary condition on both matter and
EM fields. In this case if we redo the calculations, we conclude
that this situation leads to the same result as case {\bf i}. This
shows that for a magnetodielectric medium the conditions {\bf i}
and {\bf iii} lead to the same result for the Casimir force.

\section{Conclusion}\label{conclusion}
In this article we quantized the electromagnetic field in the
presence of a magnetodielectric medium in the frame work of path
integrals. For a medium with a given susceptibility, the modified
Casimir force is obtained for different boundary conditions. The
present approach can be generalized to the case of rough perfect
conductors in the presence of a general medium straightforwardly
\cite{Fardin4}.


\vspace{-.7cm}
\begin{thebibliography}{99}
\vspace{-.7cm}
\bibitem {Casimir} H. B. G. Casimir, Proc. K. Ned. Akad. Wet.  51, 793 (1948).

\bibitem {Milonni} P. W. Milonni, R. J. Cook and M. E. Goggin, Phys. Rev. {\bf{A}} 38,  1621
(1988).

\bibitem{sriva}
Y. Srivastava, A. Widom, and M. H. Friedman, Phys. Rev. Lett. {\bf
55}, 2246 (1985); M. A. Stroscio, {\it ibid.} {\bf 56}, 2107
(1986); I. Brevik, Phys. Rev. D {\bf 36}, 1951 (1987).

\bibitem{nanoscale1}
F. M. Serry, D. Walliser, and G. J. Maclay, J. Microelectromech.
Syst. {\bf 4}, 193 (1995); J. Appl. Phys. {\bf 84}, 2501 (1998).

\bibitem{nanoscale1b}
E. Buks and M. L. Roukes, Phys. Rev. B {\bf 63}, 033402 (2001);
Nature (London) {\bf 419}, 119 (2002).

\bibitem{s2}
J. B\'arcenas, L. Reyes, and R. Esquivel-Sirvent, Appl. Phys.
Lett. {\bf 87}, 263106 (2005).

\bibitem{nanoscale2}
H. B. Chan, V. A. Aksyuk, R. N. Kleiman, D. J. Bishop, and F.
Capasso, Science {\bf 291}, 1941 (2001); Phys. Rev. Lett. {\bf
87}, 211801 (2001).

\bibitem{pinto}
F. Pinto, Phys. Rev. B {\bf 60}, 14740 (1999).

\bibitem{Cole}
D. C. Cole and H. E. Puthoff, Phys. Rev. E {\bf 48}, 1562 (1993).

\bibitem{Forward}
R. L. Forward, Phys. Rev. B {\bf 30}, 1700 (1984).

\bibitem{Lamoreaux}
S. K. Lamoreaux, Phys. Rev. Lett. {\bf 78}, 5 (1997).

\bibitem{Mohideen}
U. Mohideen and A. Roy, Phys. Rev. Lett. {\bf 81}, 4549 (1998).

\bibitem{Harris}
B. W. Harris, F. Chen, and U. Mohideen, Phys. Rev. A {\bf 62},
052109 (2000).

\bibitem{Bressi}
G. Bressi, G. Carugno, R. Onofrio, and G. Ruoso, Phys. Rev. Lett.
{\bf 88}, 041804 (2002).

\bibitem{Decca1}
R. S. Decca, D. L\'opez, E. Fischbach, and D. E. Krause, Phys.
Rev. Lett. {\bf 91}, 050402 (2003).

\bibitem{Decca2}
R. S. Decca, D. L\'opez, H. B. Chan, E. Fischbach, D. E. Krause,
and C. R. Jamell, Phys. Rev. Lett. {\bf 94}, 240401 (2005).

\bibitem {Bordag_Book}  M. Bordag, G. L. Klimchitskaya
, U. Mohideen and V. M. Mostepanenko, {\it Advanced in the Casimir
Effect}, Oxford University Press (2009).

\bibitem {Golestanian-PRL-1997} R. Golestanian and M. Kardar, Phys. Rev. Lett. {\bf 78}, 3421 (1997).

\bibitem {Golestanian-PRA-1998} R. Golestanian and M. Kardar, Phys. Rev. A {\bf 58}, 1713 (1998).

\bibitem{Emig-PRL-2001}
T. Emig, A. Hanke, R. Golestanian, and M. Kardar, Phys. Rev. Lett.
{\bf 87}, 260402 (2001).

\bibitem{Emig-PRA-2003}
T. Emig, A. Hanke, R. Golestanian, and M. Kardar, Phys. Rev. A
{\bf 67}, 022114 (2003).

\bibitem {jalal} J. Sarabadani and M.F. Miri, Phys. Rev. A {\bf 74}, 023801
(2006).


\bibitem{Fardin1}
F. Kheirandish and M. Amooshahi Phys. Rev. A {\bf 74}, 042102
(2006).


\bibitem{Fardin2}
M. Amooshahi and F. Kheirandish Phys. Rev. A {\bf 76}, 062103
(2007).

\bibitem{Fardin3}
F. Kheirandish and M. Soltani, Phys. Rev. A {\bf 78}, 012102
(2008).

\bibitem {Huttner}
B. Huttner and S. M. Barnett, Europhys. Lett. {\bf 18}, 487
(1992).

\bibitem {Ryder}
L. Ryder,\textit{ Quantum Field Theory}, Second ed. (Cabbridge
University Press, 1996).

\bibitem {Matloob} R. Matloob and H. Falinejad, Phys. Rev A, {\bf 64}, 042102
(2001).

\bibitem{Fardin4}
In preparation by the authors.

\bibitem{Lifshitz}
L.D. Landau and E.M. Lifshitz, {\it Statistical Physics}, Part 1,
Pergamon Press (1981).

\bibitem{Milonni-Book}
P.W. Milonni, {\it The Quantum Vacuum, An Introduction to Quantum
Electrodynamics}, Academic Press (1994).

\end{thebibliography}
\end{document}